\documentclass{article}

\usepackage[preprint]{neurips_2019}

\usepackage[utf8]{inputenc} 
\usepackage[authoryear]{natbib}
\usepackage[T1]{fontenc}    
\usepackage{hyperref}       
\usepackage{url}            
\usepackage{booktabs}       
\usepackage{amsfonts}       
\usepackage{nicefrac}       
\usepackage{microtype}      
\usepackage{mathtools}
\usepackage{caption}
\usepackage{multirow}
\usepackage{graphicx}
\usepackage[font=small]{caption}
\setcitestyle{square}
\usepackage{lineno}
\usepackage{subcaption}

\title{Chemical Reaction Extraction from Long Patent Documents\\[0.5em] }

\author{
    Aishwarya Jadhav\\
    Language Technologies Institute \\
    Carnegie Mellon University \\
    \texttt{anjadhav@cs.cmu.edu} \\
    \And
    Ritam Dutt \\
    Language Technologies Institute \\
    Carnegie Mellon University \\
    \texttt{rdutt@cs.cmu.edu} 
}
\begin{document}

\maketitle
 \begin{abstract} 
The task of searching through patent documents is crucial for chemical patent recommendation and retrieval. This can be enhanced by creating a patent knowledge base (ChemPatKB) to aid in prior art searches and to provide a platform for domain experts to explore new innovations in chemical compound synthesis and use-cases. An essential foundational component of this KB is the extraction of important reaction snippets from long patents documents which facilitates multiple downstream tasks such as reaction co-reference resolution and chemical entity role identification. In this work, we explore the problem of extracting reactions spans from chemical patents in order to create a reactions resource database. We formulate this task as a paragraph-level sequence tagging problem, where the system is required to return a sequence of paragraphs that contain a description of a reaction. We propose several approaches and modifications of the baseline models and study how different methods generalize across different domains of chemical patents.  

\end{abstract}

\section{Introduction}

The exponential publication rate in recent years and the fact that new innovations for chemical compound synthesis and use-cases are primarily mentioned first in patents before they appear in scientific articles makes the task of patent recommendation and retrieval crucial. However, publicly available patent recommendation systems are scarce; patents are mostly searched using Google Patents or USPTO, where recommendations are carried out through citation networks and topics. We intend to leverage key information about chemical processes from patents as well as publicly available external domain knowledge in order to improve retrieval and recommendation. To that end, we explore the idea of creating a patent knowledge base (ChemPatKB) to facilitate prior art search as well provide a means for domain experts to explore the KB in natural language queries. 

Some of the major components of the proposed ChemPatKB include patents, authors, assignees, reactions, chemical compounds and roles and properties of chemicals. In this project, we deal mainly with the major recipes or reactions involved in the patent since reactions are an essential component of the Chemical KB  and help focus on the more important spans within the long patent documents. The essence of the reaction extraction task involves document-level information extraction to detect the textual spans that describe or refer to chemical reactions. 

While most previous research in text mining of chemical reactions has focussed on chemical NER, there has been limited research on automatically extracting chemical reactions from patents. A chemical reaction is a process where a set of chemical compounds is transformed into another set of chemical compounds. A reaction description may include the source chemical compounds, solvents and reagents involved in the reaction, reaction conditions, and materials obtained as a result of the reaction. Once a reaction has been identified, it can be used as the input to (more complex) downstream tasks. For example, consider an event extraction system that extracts every step of a reaction as an individual event. Such downstream tasks require as input a paragraph sequence corresponding to a reaction, in a representation that preserves the order of the reaction substeps. Our end goal is to create such a database of reaction sequences, much larger and encompassing more domains than the ones available presently. 

We build on top of the baseline model \cite{yoshi} proposed by Yoshikawa et al. by introducing a BERT-based embedding module, exploring sentence vs paragraph level predictions, and replacing chemical entities with special chemical tokens for better generalization. We train the models over a manually annotated dataset made available by Chemu \cite{chemu} and test generalization over a test collection consisting of patents from Organice, Inorganic, Petrochemical, and Alcohol domains. 

Code for this project can be found on GitHub \footnote{\url{https://github.com/aishwaryajadhav/Chemical-Patent-Reaction-Extraction}}. 

\section{Literature Review}

Although there has been limited work on this topic, this research direction is quite dated. Patents are regarded as an important resource for chemical information, and a large volume of NLP research has focused on them (\cite{tseng}, \cite{fuji}, \cite{guru}). Some previous work has attempted to extract not only chemical names but also reaction procedures from the literature (\cite{lawson}, \cite{wei}, \cite{kalin}). Among them, \cite{lowe} presents an integrated system that detects reaction text from chemical patents, and extracts chemicals and their roles in the corresponding reaction. The system is heavily rule-based and incorporates existing NLP libraries. More recent efforts for the specific task of reaction extraction has been described in Yoshikawa et al., 2019 \cite{yoshi}, where a number of baseline and neural architectures have been proposed for this task. They achieve quite good results for the extraction task. However, their models were trained on a very limited set of silver standard data containing patents from organic chemistry. They do not report on how their models generalize to different topics or domains of chemical patents.

Another aspect of this problem is the availability of a standard benchmark gold dataset containing reactions extracted from chemical patents covering a variety of topics and domains. The dataset used by Yoshikawa et al. \cite{yoshi} is silver standard dataset derived from the Reaxys® database. Recently Chemu \cite{chemu} released a gold standard annotated dataset for the Chemu challenge of reaction coreference resolution resolution containing reaction span annotation for 150 patents.

In this work we try to improvise on models for reaction extraction and work towards generating a large scale patent-reactions resource for research in this area.

\section{Dataset}

We use the Chemu dataset \cite{chemu} for training and evaluation of our models. It consist mainly of organic chemistry patents from the European Patent Office and the United States Patent and Trademark Office (USPTO).  It is a gold standard dataset manually annotated for the Chemu 2021 challenge \cite{chemu}. 

The dataset contains 120 annotated patent documents in the BRAT file format in the train set and 30 patents in the development set. The annotations are available in the form of character-level spans denoting the reactions. Additional annotations include relations between parent-child reactions and relation identification cues which are ignored for our task. Table \ref{data-stats} contains some stats about the patent documents and reaction annotations. We use the 30 dev set documents as a blind test set and use 20\% of the 120 train set documents as the validation set. The rest is used for training.

\begin{table}[h]
\centering
\resizebox{0.4\columnwidth}{!}{
\begin{tabular}{|l|l|l|}
\hline
& Train Set & Dev Set \\
    \hline
   No of Files & 120 & 30 \\
No of Words & 1186K & 295K \\
No of Paras & 	53.5K	& 12.5K \\
Avg No of tokens per para &	22.15	& 22.9\\
Avg No of paras per document & 446.4	& 429 \\
Total reactions annotated &	6378	& 1244 \\
    \hline

\end{tabular}
}
\caption{Gold Dataset Statistics}
\label{data-stats}
\end{table}

We also test the generalization performance of our models on out-of-domain patents. For this, we have hand-picked a collection of 4 organic chemistry patents belonging to various CPC codes and a set of 3 patents from inorganic chemistry, petrochemical and alcohol domains. All of these contain varying citations, are from industry or academia and small and large corporations to add variability to the style to evaluate the robustness of our models. This formed our generalization set.

\section{Task Formulation}
Multiple contiguous paragraphs often describe a single reaction. Since a reaction consists of a series of sub-steps executed over time, it is crucial to accurately detect the beginning and end of each reaction text. Therefore, we define the task as a span detection problem rather than the simpler task of binary classification (i.e., classifying each paragraph as describing (part of) a reaction or not) to capture reaction substructure. A patent document is given as a sequence of paragraphs. The task is to detect a span of contiguous paragraphs that describe a single chemical reaction. In our corpus, we provide paragraph-level label sequences over paragraphs in patent documents, following the IOB2 tagging scheme \cite{iob}.

We transform the character span annotations to IOB tags using a simple mapping that assigns a ‘B’ tag to the paragraphs containing the first character of the reaction span and ‘I’ tag to all the subsequent paragraphs uptil the one containing the last reaction span character. All other paragraphs that do not contain any reaction characters are tagged ‘O’.

\section{Baseline Models}

\begin{figure*}
\begin{center}
\includegraphics[width=13cm]{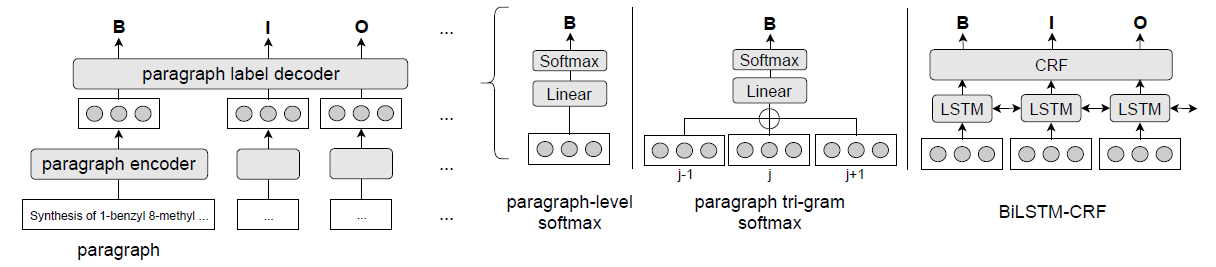}
\caption{Baseline Architecture. The left figure illustrates the general architecture of the whole model, while the
right figure details the 3 decoder components. }
\label{baseline}
\end{center}
\end{figure*}

Our backbone architecture is based on the baseline models described in Yoshikawa et al. \cite{yoshi}. Figure \ref{baseline} from the paper describes these architectures. All 3 baseline models have the same paragraph encoders but use different decoders for generating the IOB tags of a sequence of patent paragraphs. 

On a high level, each paragraph is tokenized into words using the OSCAR4 tokenizer \cite{jessop}, which is specifically customized for chemical text mining. The embedding for each token is then generated by concatenating the pre-trained word embedding of each token or word $wi$: $e^{WE}_{w_i}$, its contextualized embedding $e^{CW}_{w_i|p}$, and an optional embedding $e^{FT}_{fi}$ representing additional features $f_i$ associated with $w_i$: 
$$ e_i = e^{WE}_{w_i} + e^{CW}_{w_i|p} + e^{FT}_{fi}$$
For the word embeddings, $e^{WE}_{w_i}$ and contextualized embeddings $e^{CW}_{w_i|p}$, Word2Vec \cite{miko} and ELMo \cite{peters}, respectively, are employed. Both are pre-trained on chemical patent documents from \cite{zhai}. These embeddings are fixed during training. The additional learnable feature embeddings $e^{FT}_{fi}$ (in Equation 1) are based on the output of a chemical named entity recognizer \cite{zhai}. 

The paragraph encoder then generates a single embedding vector for the entire paragraph using a BiLSTM over the concatenated token embeddings. The paragraph decoders take in the para embeddings to output a B/I/O tag for each paragraph. The following 3 decoder models are used for the 3 baselines as shown in figure \ref{baseline}. The linear softmax decoder considers just the current paragraph while predicting the B/I/O tags. The Trigram-based decoder considers one paragraph before and one after the paragraph being decoded and tagged. The BiLSTM-CRF based decoders helps capture long range dependencies between the paragraphs.

\section{Our Approach}

We utilize the paragraph encoders described in [1]. For the decoder, we experiment with both the trigram and BiLSTM CRF-based architectures. We did run 1 experiment with the linear-softmax based decoder using our approach described below but choose not to use it in other experiments due to its sub-optimal results.

\subsection{BERT Embeddings}

For the 2 baseline models described above, we replace the first embedding layer that feeds into the paragraph encoder by BERT-based embeddings. BERT is a SOTA model used for a wide range of NLP and IR tasks and we wanted to assess the performance of the baseline architecture with BERT embeddings. We conducted experiments using raw token embeddings generated without finetuning the base BERT (bert-base-uncased), with embeddings generated by finetuning base BERT (bert-base-uncased).

\subsection{ChemBERT}
We also experimented with 2 different pretrained BERT models:

\begin{itemize}
\item \textbf{recobo/chemical-bert-uncased-pharmaceutical-chemical-classifier}: Chemical domain BERT model finetuned on 13K Chemical, and 14K Pharma Wikipedia articles broken into paragraphs.
\item \textbf{recobo/chemical-bert-uncased}:
BERT-based model further pre-trained from the checkpoint of SciBERT using a corpus of over 40,000+ technical documents from the Chemical Industrial domain and combined it with 13,000 Wikipedia Chemistry articles, ranging from Safety Data Sheets and Products Information Documents.
\end{itemize}

\subsection{Sentence-level Sequence Tagging}

During our experiments, we observed that most of the errors were witnessed at the end boundary of the reaction, especially where properties of the newly synthesized compound were being listed. Some patents contained these properties in the same paragraph as the reaction paragraphs while others mentioned them in a separate paragraph. So the models always made errors while determining the correct end boundary for the reaction span. This was actually an artifact of our ground truth tagging scheme. Because we formulated the problem as a paragraph-level sequence tagging task, we consider the entire paragraph as a reaction even though a part of it is outside the actual annotated character-level reaction span. About 8\% of the reaction paragraphs in the train set contained more 40\% non-reaction characters.

Hence, we experimented with sentence-level encoding and tagging. The architectures here are the same as that of the paragraph-level tagging task. Instead, the training and validation data contain tags at the sentence level.

\subsection{[CHEM] tokens}

When we checked the generalization performance of the models outlined above on the out-of-domain generalization set, we observed that most of the mistakes occurred where the reaction paragraph described a reaction without any chemical names. This could be paragraphs describing general reaction recipes or that the domain addresses its materials by common language names instead of their chemical nomenclature. In order to combat the model overfitting to chemical compound names, we replace all chemical names in the train, validation, test and generalization set with a new token [CHEM]. This was added to the BERT layer as a special token. We then finetuned the bert-base-uncased and chemical-bert-uncased-pharmaceutical-chemical-classifier with this new token. 

In order to tag the chemical tokens in the patents to replace with [CHEM], we used tmChem \cite{tmchem} an open-source software tool for identifying chemical names in biomedical literature, including chemical identifiers, drug brand and trade names. This tagger service was accessed over a RESTful api made available by the authors that parses the entire patent document or chucks thereof in a single request returning all the chemical name tokens in the submitted text. 

\section{Evaluation Metrics}

For model selection we use the span-based scores based on a strict match strategy, where an output span is regarded as correct if the beginning and ending paragraphs strictly match those of the gold span. In some practical applications, it also makes sense to understand if the model can identify the approximate region where a reaction is described. Thus, for evaluation, we also compute the scores based on a fuzzy match strategy, where we calculate the number of matches by counting the number of gold spans that have at least one corresponding predicted output span whose beginning and ending paragraph indices are at most 1 paragraph away from the gold ones.

We report the Accuracy, Precision, Recall and F1 score for each model on the test and generalization datasets according to the strict and fuzzy match criteria defined above.

\section{Results and Error Analysis}

We first evaluate the performance of the various approaches over the in-domain test set. Note that we do not show the results of the sentence based models since these results on the test and generalization are extremely bad. Sentence based tagging needs the model to capture very long range dependencies because a single reactions paragraph can contain tens of sentences. Hence, capturing multi-para reactions necessitates attending to a long sequence of reaction sentences which does not work well at all.
\subsection{Performance on In-Domain Test Set}

\subsubsection{Trigram vs BiLSTM CRF decoder for base BERT models}

\begin{table}[h]
\centering
\resizebox{0.99\columnwidth}{!}{
\begin{tabular}{|l|l|l|l|l|l|l|}
\hline
\multirow{3}{*}{Model} & 
    \multicolumn{3}{c}{Strict Match} &
    \multicolumn{3}{c}{Fuzzy Match} \\
    & F1 & Precision & Recall & F1 & Precision & Recall \\
    \hline
    Trigram Base BERT & 0.727 & 0.749	&  0.706	& 0.813	& 0.799	& 0.827 \\
    Trigram Base BERT Finetuned & 0.714 &	0.769 &		0.667 &		0.836 &		0.836	 &	0.836 \\
    BiLSTM CRF Base BERT & 0.678 &	0.713 &	0.646 &	0.825 &	0.841 &	0.81 \\
    BiLSTM CRF Base BERT Finetuned & 0.745 &	0.8135 &	0.6875	 & 0.853 &	0.8975 &	0.814\\
    \hline

\end{tabular}
}
\caption{Trigram vs BiLSTM CRF decoder for base BERT models}
\label{tri-crf}
\end{table}

Table \ref{tri-crf} depicts the comparison between the Trigram-based decoder vs the BiLSTM CRF-based decoder for the bert-base-uncased BERT embeddings. We compare the performances of the models with and without BERT finetuning. The fuzzy f1 score for the CRF-based models are higher in both the scenarios. We use 16 paragraph sequences for the BiLSTM encoding-decoding and CRF decoding. Naturally, the CRF model is better able to capture the long range dependencies as compared to the trigram decoder model that just considers 1 paragraph before and after the current paragraph. Hence, for the future studies, we only train and evaluate the BiLSTM-CRF based architecture.

We also observe that fine-tuning the BERT model provides significant gains in both scenarios. 

\subsubsection{Base BERT vs ChemBERT models}

\begin{table}[h]
\centering
\resizebox{0.99\columnwidth}{!}{
\begin{tabular}{|l|l|l|l|l|l|l|}
\hline
\multirow{3}{*}{Model} & 
    \multicolumn{3}{c}{Strict Match} &
    \multicolumn{3}{c}{Fuzzy Match} \\
    & F1 & Precision & Recall & F1 & Precision & Recall \\
    \hline
    Base BERT Finetuned & 0.745 &	0.8135 &	0.6875	 & 0.853 &	0.8975 &	0.814\\
    
    ChemBERT No Finetuning  & 0.712 & 	0.706	 & 0.717179903	 & 0.819 & 	0.792	 & 0.848
 \\
    ChemBERT Finetuned & 0.727 & 	0.758 & 	 0.699	 & 0.854 & 	0.855	 & 0.853
 \\
    \hline

\end{tabular}
}
\caption{BiLSTM CRF models with Base BERT vs ChemBERT embeddings}
\label{base-chem}
\end{table}

Table \ref{base-chem} summarizes the performances obtained by using the BERT pretrained on the chemical domain dataset for generating the token embeddings against the Base BERT performance. We note that the ChemBert used here is the recobo/chemical-bert-uncased-pharmaceutical-chemical-classifier pretrained BERT since this gave superior results as compared to the other pretrained BERT model. Hence, we only report results and perform comparative analysis for this version of chemical pretrained BERT.
Here again we see that finetuning benefits the performance considerably. We observe minor gains in the performance of the model based on ChemBert embeddings. This improvement is not significant. Although we would have expected a good improvement upon using the ChemBERT embeddings, model finetuning closes this gap.

\subsubsection{Effect of Introducing [CHEM] tokens}

\begin{table}[h]
\centering
\resizebox{0.99\columnwidth}{!}{
\begin{tabular}{|l|l|l|l|l|l|l|}
\hline
\multirow{3}{*}{Model} & 
    \multicolumn{3}{c}{Strict Match} &
    \multicolumn{3}{c}{Fuzzy Match} \\
    & F1 & Precision & Recall & F1 & Precision & Recall \\
    \hline
    Base BERT No Finetuning ([CHEM]) & 0.729 & 	0.739	 & 0.72 & 	0.839	 & 0.83 & 	0.848 \\
    Base BERT Finetuned ([CHEM]) & 0.72 & 	0.799 & 	0.655 & 	0.844 & 	0.887 & 	0.806 \\
    ChemBERT No Finetuning ([CHEM])  & 0.643 & 	0.685 & 	0.607	 & 0.819	 & 0.846	 & 0.793 \\
    ChemBERT Finetuned ([CHEM])  & 0.742 & 	0.777 & 	0.709 & 	0.833	 & 0.848 & 	0.819  \\
    \hline

\end{tabular}
}
\caption{BiLSTM CRF models with [CHEM] tokens}
\label{chem-token}
\end{table}

The main observation in this scenario is that unlike the previous case, here, the base BERT model outperforms the ChemBERT-based model. However, this is not surprising because ChemBERT has been trained to recognize and embed chemical names and tokens which are masked in this scenario. Hence, using ChemBERT in the absence of any chemical entity tokens hurts the model. We also observe that the [CHEM] masked model with base bert performs similarly to the non-CHEM masked model from table \ref{chem-token}. This hints toward the fact that the model actually learns the structure of the reaction paragraphs rather than basing the reaction or no-reaction decision on the presence or absence of certain chemical entities which might be different across different domains.

We now look at the generationalization performances of some models selected from the analysis above.

\subsection{Generalization Performance Analysis}

\begin{table}[h]
\centering
\resizebox{0.99\columnwidth}{!}{
\begin{tabular}{|l|l|l|l|l|l|l|}
\hline
\multirow{3}{*}{Model} & 
    \multicolumn{3}{c}{Strict Match} &
    \multicolumn{3}{c}{Fuzzy Match} \\
    & F1 & Precision & Recall & F1 & Precision & Recall \\
    \hline
    Base BERT Finetuned & 0.707 & 	0.854 & 	0.603	 & 0.776	 & 0.951	 & 0.655 \\
    Base BERT Finetuned ([CHEM]) & 0.638 & 	0.833 & 	0.517	 & 0.775	 & 0.949 & 	0.655 \\
    ChemBERT Finetuned & 0.708 & 	0.769 & 	0.656 & 	0.801	 & 0.855 & 0.754 \\
    ChemBERT Finetuned ([CHEM]) & 0.705 & 	0.787	 & 0.638	 & 0.850	 & 0.94	 & 0.776 \\
    \hline

\end{tabular}
}
\caption{Generalization Performances of BiLSTM CRF Models}
\label{gen}
\end{table}

While the strict match performance of all the models here is the same, we see that the results for the fuzzy match are quite the opposite of those seen for the test set. The ChemBERT-based models significantly outperform the base BERT models. Even for the approach consisting of [CHEM] tokens, the ChemBERT based embeddings provide much better results. One plausible explanation for this might be the fact that this chemical domain pretrained model is actually learning the structural components of  reaction paragraphs which are similar across domains. Just the entities in the reactions are different for the different domains and thus, replacing those with [CHEM] tags reduces the noise and enables the model to focus on the structure of the paragraph to determine whether it is a reaction paragraph or not. 

A main problem observed across the different models for the generalization set is the poor recall of these models. Both the ChemBERT models have also performed significantly better in terms of extracting more reaction spans present in the patents, demonstrating generalizability across different domain chemical patents.

\section{Discussion and Future Direction}

In this project, we primarily explored various methods to extract reaction spans from patents and evaluated their performance in both in-domain and out-of-domain chemical patent documents. Our findings highlight several areas where current models struggle. Firstly, the models often perform poorly for reaction spans exceeding three paragraphs. Secondly, they encounter difficulties in accurately demarcating boundaries between consecutive reactions, especially when multiple reactions follow one another without clear headings. Additionally, the models are not well-equipped to handle reaction snippets embedded within tables.

Moving forward, our research will focus on two main directions. Firstly, we plan to explore the impact of multi-task learning on reaction extraction. Specifically, we aim to evaluate the performance of a BERT-based model trained jointly for Chemical Named Entity Recognition (NER) and reaction extraction. This approach will help us determine whether learning structural components of the text through joint training for these two tasks enhances the model's ability to generalize across domains. Secondly, we intend to create a comprehensive database of reactions extracted from chemical patents for use in downstream tasks.

Moreover, there is a pressing need for a standardized benchmark dataset for the Reaction Extraction task. Although our results demonstrate superior performance compared to previous work by Yoshikawa et al., we emphasize the importance of a gold standard dataset with annotations for a broader set of patents. Such a benchmark is crucial for fair comparisons across different methodologies in this area.

\section{Conclusion }
We have explored different approaches to extracting reaction spans from chemical patent documents and evaluated the generalization of these models across various chemical domains. We seek to continue to improve this task in order to create a valuable resource of reactions for the research community.

\bibliography{chem_extract}
\clearpage

\end{document}